\documentclass{aa}
\usepackage{txfonts}
\usepackage[utf8]{inputenc}
\usepackage{graphicx}
\usepackage{natbib}
\usepackage{amssymb}
\usepackage{amsmath}
\usepackage{booktabs}
\usepackage{pdfpages}
\usepackage{dblfloatfix}
\usepackage[colorlinks=true,citecolor=blue,linkcolor=blue,urlcolor=blue]{hyperref}

\makeatletter
\renewcommand*\aa@pageof{, page \thepage{} of \pageref*{LastPage}}
\makeatother

\newcommand{\jxs}{\texttt{JoXSZ}}
\newcommand{\mb}{\texttt{MBProj2}}
\newcommand{\ppf}{\texttt{PreProFit}}
\newcommand{\clj}{CL~J1226.9+3332}

\begin{document}

\title{JoXSZ: Joint X-SZ fitting code for galaxy clusters}
\author{Fabio Castagna\inst{1} \and Stefano Andreon\inst{1} 
}
\institute{INAF–Osservatorio Astronomico di Brera, via Brera 28, 20121 Milano, Italy \\ \email{fabio.castagna@inaf.it} 
}

\abstract{The thermal Sunyaev-Zeldovich (SZ) effect and the X-ray emission offer separate and highly complementary probes of the thermodynamics of the intracluster medium.
We present \jxs, the first publicly available code designed to jointly fit SZ and X-ray data coming from various instruments to derive the thermodynamic profiles  of galaxy clusters. \jxs\  follows a fully Bayesian forward-modelling approach, accounts for the SZ calibration uncertainty, and for the X-ray background level systematic. It improves upon most current and not publicly available analyses because it adopts the correct Poisson-Gauss expression for the joint likelihood, makes full use of the information contained in the observations, even in the case of missing values within the datasets, has a more inclusive error budget, and adopts a consistent temperature in the various parts of the code, allowing for differences between X-ray and SZ gas-mass weighted temperatures when required by the user.
\jxs\  accounts for beam smearing and data analysis transfer function, accounts for the temperature and metallicity dependencies of the SZ and X-ray conversion factors, adopts flexible parametrisation for the thermodynamic profiles, and on user request, allows either adopting or relaxing the assumption of hydrostatic equilibrium (HE).
When HE holds, \jxs\  uses a physical (positive) prior on the radial derivative of the enclosed mass and derives the mass profile and overdensity radii $r_\Delta$. For these reasons, \jxs\  goes beyond simple SZ and electron density fits. 
We illustrate the use of \jxs\  by combining Chandra and NIKA data of the high-redshift cluster \clj.
The code is written in Python, it is fully documented, and the users are free to customise their analysis in accordance with their needs and requirements. \jxs \ is publicly available on GitHub.
}
\maketitle

\section{Introduction} \label{sec:intro}
Galaxy clusters trace the backbone of the Universe, and their thermodynamic properties are valuable assets, for instance, for probing the physical structure and substructure of the intracluster medium (ICM) to unveil the occurrence of phenomena such as merger events \citep{Mroczkowski2019}. For example, the cluster thermal history is fully captured by the entropy, and large-scale deviations from a power-law behaviour can be used to examine how the ICM is affected by non-gravitational processes, such as heating and radiative cooling of active galactic nuclei (AGN) \citep{Voit2005a}. The cooling-time profile may be used to distinguish between clusters with and without a cool core \citep{Hudson2010}, while the gas fraction directly relates to the strength of radiative cooling and star formation \citep{Sun2009}. Sharp jumps in the temperature or pressure profiles generally indicate shocks and cold fronts \citep{Markevitch2007}.
Furthermore, thermodynamic profiles allow us to infer the mass distribution inside the cluster, and from the latter, to improve cosmological constraints \citep{Bocquet2015, Ruppin2019c}, such as the dark energy equation of state ($w$), the number of neutrino species, the matter density ($\Omega_M$), and the amplitude of matter power spectral fluctuations on 8 Mpc h$^{-1}$ scales ($\sigma_8$).

Thermodynamic profiles of galaxy clusters can be derived from observations in the optical band, the X-ray band, or in microwaves in the shape of the Sunyaev-Zeldovich (SZ) effect \citep{Sunyaev1970, Sunyaev1972}. 
In particular, SZ measurements became widespread in the past decade \citep[e.g.][]{Birkinshaw2005, Mroczkowski2009, Korngut2011, Sayers2013, Adam2015, Romero2017} because of the huge increment in high-resolution SZ instruments \citep[see][for details]{Mroczkowski2019, Castagna2019}.

Combining SZ and X-ray observations to gather thermodynamic profile is extremely advantageous because the two wavelengths encode information about the ICM \citep{Ameglio2007}. 
The highest benefit of this approach applies to high redshift and to the outskirts of the cluster, where the very low surface brightness of the X-ray signal and its low contrast against the background leads to noisy temperature estimates that are often affected by systematics related to the difficult X-ray background spectral modelling \citep[e.g.][]{Eckert2011}. On the other hand, SZ measurements do not experience dimming because they are nearly independent of redshift. However, the joint derivation method is not straightforward, mostly because of cross-correlation among thermodynamic measures: in SZ, the conversion factor from Compton $y$ to surface brightness is temperature dependent (a change within 5\% between $T=5$ keV and $T=10$ keV at 150 GHz, within 15\% at 260 GHz), and in X-ray, the conversion from electron density to a soft-band count rate depends on both temperature (by 2\% for the same temperature change) and metallicity \citep[20\% change from 0.3 solar to 1 solar;][]{Ettori2013}. Furthermore, wavelength-specific temperatures may differ from one another: the SZ temperature is gas-mass weighted, while the X-ray temperature is derived spectroscopically.

Several authors have paved the way for joint X-SZ analyses, mostly considering electron density and SZ data \citep[e.g.][]{Kitayama2004, Ameglio2007, Mroczkowski2009, Eckert2013, Adam2015, Shitanishi2018, Siegel2018, Ghirardini2018, Ruppin2019b}, where electron density and SZ strength were often derived for a given temperature and metallicity.
In a different approach, we replace the electron density fit with a full spatial-spectral X-ray data fit. This provides consistent and improved estimates of all thermodynamic profiles, and the temperature and metallicity dependencies are included in the error budget.

This paper presents \jxs, the first publicly available program for performing a multiwavelength joint fit of SZ and X-ray data on galaxy clusters.
\jxs\  supports data coming from different X-ray or SZ telescopes, follows a fully Bayesian forward-modelling approach, and supports flexible modelling of the thermodynamic components of a cluster.
Users are free to choose which parameter to fit, to decide whether hydrostatic equilibrium should be adopted, whether there are systematics between X-ray and SZ temperatures, and other options. \jxs\  returns the maximised likelihood, model estimates with uncertainties, joint and marginal probability contours, convergence diagnostics, and projected radial profiles, as detailed below. When data are missing in the input file, for instance, measurements are lacking at some radii, \jxs \ is able to deal with them automatically by excluding values marked as missing (nan) from the likelihood computation.

The paper is organised as follows: in Sect.~\ref{sec:jxs} we provide an overview of the software and the technical requirements; in Sect.~\ref{sec:methods} we describe in detail the method behind each step of the program; in Sect.~\ref{sec:ex} we present an application of \jxs \ to real data from the galaxy cluster \clj; and we conclude with the discussion and final remarks in Sect.~\ref{sec:conc}. Appendix~\ref{app:py} presents the technical implementation of \jxs.

\section{\jxs} \label{sec:jxs}

\subsection{Program flow}
\begin{figure}[t!]
    \centering
    \includegraphics[width=\linewidth]{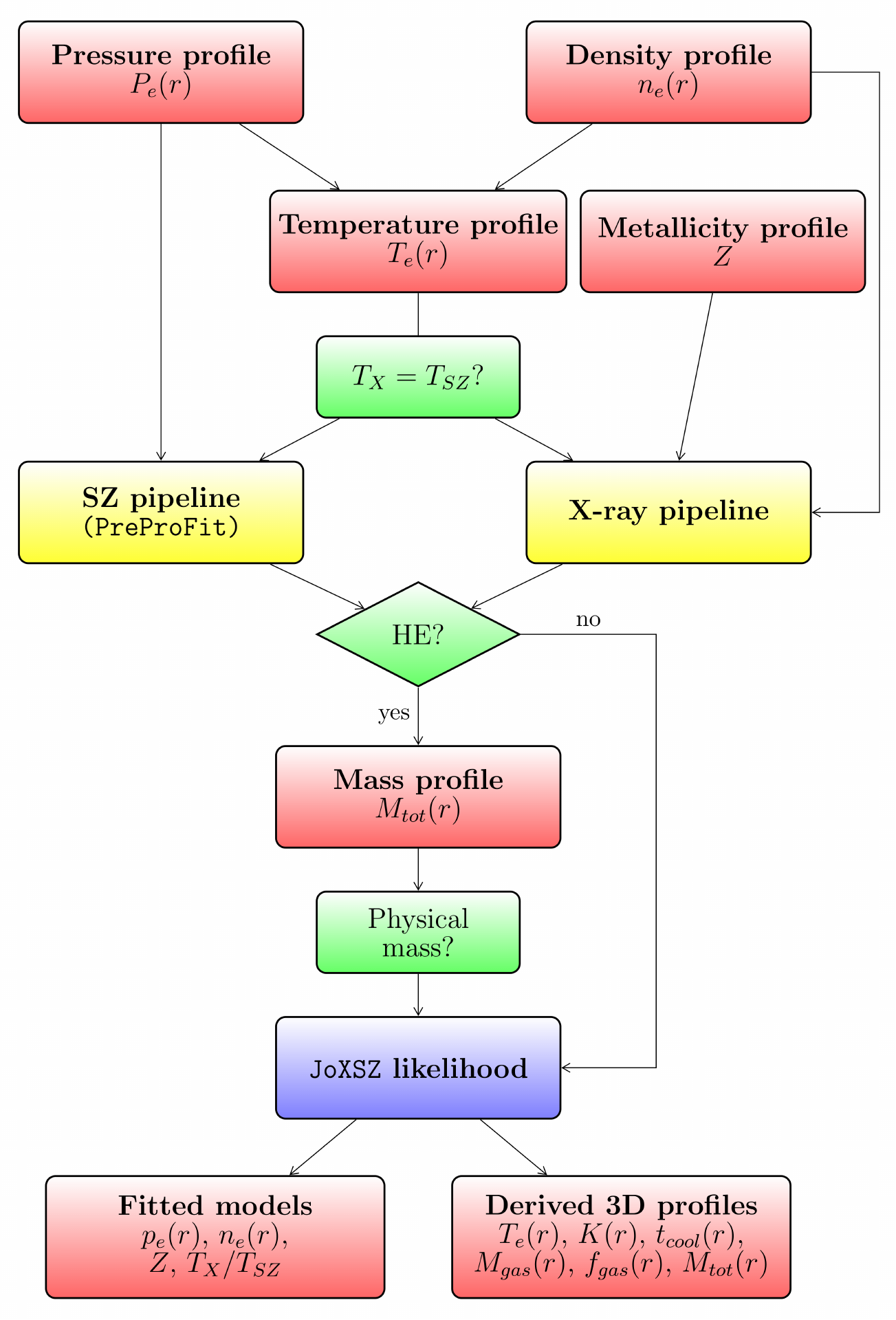}
    \caption{Block diagram showing the program flow. Radial profiles are plotted in red. Options are shown in green. Analysis pipelines are depicted in yellow. Data enter in the blue box.}
    \label{fig:flow}
\end{figure}

As represented in Fig.~\ref{fig:flow}, \jxs\  adopts a descriptive model for a spherically symmetric galaxy cluster with a given centre that parametrises its pressure and electron density profile, and derives the temperature profile as the ratio of these quantities through the ideal gas law.  The X-ray and SZ-based temperatures can be asked to be consistent or to differ, for example to study the cluster elongation along the line of sight, gas clumping, or calibration uncertainties.
The model assumes a flat metallicity profile for the cluster, whose value can be fitted, or fixed, at user request. 
Other thermodynamic quantities (entropy, cooling time, and gas mass) are automatically computed, output, and plotted by \jxs. \jxs\  fits the surface brightness profile in the SZ domain and across multiple X-ray bands. 
Under the HE assumption, \jxs\  also computes the mass profile and the gas fraction profile, and derives $r_\Delta$ and $M_\Delta$. 
The SZ and X-ray modelling structures rely on the \ppf \citep{Castagna2019} and \mb \citep{Sanders2018} pipelines, respectively. \jxs\  merges these two processes into a unique joint and consistent model based on a Markov chain Monte Carlo (MCMC) fitting algorithm.

\subsection{Requirements and installation}
\jxs\  was developed and tested with Python 3.6. The following libraries are required: mbproj2, PyAbel, numpy, scipy, astropy, emcee, six, matplotlib, and corner. 
\jxs\  can be downloaded from GitHub\footnote{\url{https://github.com/fcastagna/JoXSZ}}.

\section{Methods} \label{sec:methods}

\subsection{Intracluster medium modelling} \label{subsec:parametrization}

\subsubsection{Pressure profile} \label{subsec:press}
The pressure profile is described by the generalised Navarro, Frenk \& White (gNFW) model proposed by \cite{Nagai2007},
\begin{equation}
    P_e(r)=\frac{P_0}{\left(\frac{r}{r_p}\right)^c\left(1+\left(\frac{r}{r_p}\right)^a\right)^{\frac{b-c}{a}}},
    \label{eq:press_prof}
\end{equation}
where $P_0$ is a normalising constant and $r_p$ is a scale radius. The exponentials $b$ and $c$ describe the logarithmic slopes at $r/r_p\gg1$ and $r/r_p\ll1$, respectively, while $a$ governs the turnover rate between these two slopes. The five parameters mean that the model is very flexible in fitting current data.

The three-dimensional pressure model is numerically integrated along the line of sight in order to obtain the two-dimensional map of the Compton $y$ parameter, built on a regular grid assuming radial symmetry. 
As discussed in detail in \citet{Castagna2019}, the integration is conducted via Abel transform and the upper integration limit $R_b$ is at user choice.

\subsubsection{Density profile}
To parametrise the electronic density profile, we adopted the model introduced by \cite{Vikhlinin2006},
\begin{equation}
    n^2_e(r)=\frac{n^2_{e0}}{\left(\frac{r}{r_c}\right)^{\alpha}\left[1+\left(\frac{r}{r_c}\right)^2\right]^{3\beta-\frac{\alpha}{2}}\left[1+\left(\frac{r}{r_s}\right)^\gamma\right]^{\frac{\epsilon}{\gamma}}}+\frac{n^2_{e02}}{\left[1+\left(\frac{r}{r_{c2}}\right)^2\right]^{3\beta_2}},
    \label{eq:dens}
\end{equation}
where $n_{e0}$ is a normalising constant, $\alpha$ is the logarithmic slope at $r/r_c \ll 1$, $r_c$ represents the core radius, $\beta$ is the shape parameter for the isothermal $\beta$-model \citep{Cavaliere1976}, $r_s$ is the radius at which the density profile steepens with respect to the traditional $\beta$-model, $\epsilon$ interprets the change in slope near $r_s$ , and $\gamma$ measures the width of the transition region.
The second $\beta$ component, that is, the additive term in Eq.~\ref{eq:dens}, increases model flexibility close to the centre of the cluster by including a small core radius $r_{c2}$ with shape parameter $\beta_2$ and additive constant $n_{e02}$.

\subsubsection{Temperature profile} \label{subsec:temp}
Assuming the ideal gas law, we derive the temperature profile as the ratio between the pressure and density profiles,
\begin{equation}
    k_BT_{e,SZ}(r)=\frac{P_{e,SZ}(r)}{n_{e,X}(r)},
    \label{eq:temp_profile}
\end{equation}
where $k_B$ is the Boltzmann constant.
This temperature calculation is simultaneously influenced by the pressure constraints from SZ observations and the electron density constraints from X-ray observations.
Temperature gradients manifest themselves as gradients in the ratio of X-ray surface brightness profiles in different bands.
As already mentioned and as detailed in the diagram in Fig.~\ref{fig:flow}, \jxs \ allows users either to consider a unique temperature profile $T_{SZ}=T_X$, or to make a distinction between the gas-mass weighted temperature $T_{SZ}$ and the X-ray temperature $T_X$, introducing the multiplicative parameter $\log(T_X/T_{SZ})$.
Thus, \jxs\  assumes identical shapes for the two temperature profiles, only allowing different normalisations.
The profiles of the entropy $K(r)$, cooling time $t_{cool}(r)$, and gas mass $M_{gas}(<r)$ are computed as usual \citep{Sanders2018}.

\subsubsection{Mass distribution} \label{subsec:mass}
If requested, the total mass distribution can be derived assuming hydrostatic equilibrium (HE).
The mass $M(<r)$ enclosed within a radius $r$ is related to the pressure and electronic density profiles through
\begin{equation}
    M_{tot}\left(<r\right)=-\frac{r^2}{\mu_{gas}m_pGn_e(r)}\frac{dP_e(r)}{dr},
    \label{eq:HE}
\end{equation}
where $\mu_{gas}=0.61$ is the mean molecular gas mass \citep{Anders1989}, $m_p$ is the proton mass, and $G$ is Newton's constant. 
Because the total mass distribution as defined in Eq.~\ref{eq:HE} directly depends on the pressure and electron density parameters, a non-monotonically increasing mass profile may occur, and this means that the program would assign negative values of mass at some radii. To avoid this, a positive prior on the radial derivative of the enclosed mass is imposed by default, although users can remove it.

The adopted version of the hydrostatic equilibrium formula exploits the advantages of SZ and X-ray surface brightness profiles over X-ray temperature estimates, especially
at large distances from the cluster centre. \jxs\  uses the analytic expression for the pressure derivative to improve program execution time. 

By comparing the mass distribution in Eq.~\ref{eq:HE} with its definition in terms of volume and density,
\begin{equation}
    M\left(<r_\Delta\right)=\frac{4}{3}\pi\rho_c(z)\Delta r_\Delta^3,
\end{equation}
the overdensity radius $r_\Delta$ is derived as the radius within which the average density is $\Delta$ times the critical density at the cluster redshift, $\rho_c(z)$. $M_\Delta$ computation ensues as $M_\Delta=M\left(r_\Delta\right)$.

The gas fraction profile is straightforwardly deduced as
\begin{equation}
    f_{gas}\left(r\right)=\frac{M_{gas}\left(<r\right)}{M_{tot}\left(<r\right)},
\end{equation}
where $M_{gas}\left(<r\right)$ is obtained by integrating the product between the gas density profile and the volume of the shell.

\subsubsection{Conversions from physical to instrumental quantities} \label{subsec:conv}
In X-ray, the conversion from emissivity to count-rate depends on the gas temperature $T_X$, metallicity, and Galactic absorption.
As in \citet{Sanders2018}, \jxs\  uses XSPEC \citep{Arnaud1996} to compute this conversion, which requires users to have computed the response matrix file (RMF) and the ancillary response file (ARF).

In SZ, the conversion from Compton $y$ to surface brightness depends on the temperature $T_{SZ}$ because of the relativistic corrections to the SZ effect \citep{Itoh2004}. The exact value depends on the specific response of the instrument, and thus users are required to provide the instrument-specific conversion factor as an input file. \jxs\  also accounts for the SZ calibration uncertainty, implemented as a Gaussian whose parameters can be set by the user. Both SZ and X-ray conversion factors are depend on the radius and are updated at each
step of the chain.

\subsection{Processing backbones: \ppf\  and \mb} \label{subsec:backbones}
\jxs\  relies on \ppf\  \citep{Castagna2019} and \mb\  \citep{Sanders2018}, where full details on the SZ and X-ray processing are presented.
In a nutshell, \ppf \ takes as input the pressure profile, whose sampling step is at user choice, then projects it onto a two-dimensional map using forward Abel transform, convolves the map with the instrumental beam and the transfer function, and finally derives the surface brightness profile through opportune conversion factors.

About the X-ray analysis, \mb\  makes full use of the spatial-spectral X-ray data cube and allows users to model one or more thermodynamic profiles, but not the pressure profile.
\jxs \ instead parametrises it together with the electron density profile and derives the remaining ones from the ideal gas law (see Appendix~\ref{app:py} for the technical implementation in Python). As mentioned, with the further assumption of hydrostatic equilibrium, \jxs \ computes the mass profile and the characteristic radius $r_{\Delta}$. The X-ray fitting procedure includes a component that accounts for the background level systematic. As anticipated in Sect.~\ref{subsec:temp}, the user is free to choose whether to include a single temperature profile that is simultaneously fitted from SZ and X-ray data or to consider two separate temperature profiles.

\subsection{Model definition} \label{subsec:model}
To set up our joint model, we defined the likelihood function as the product of the two distinct functions for the SZ and X-ray analyses: the former is a $\chi^2$ function, the latter is a Poisson likelihood. As a result, the log-likelihood function $\mathrm{ln}(\mathcal{L}_{\jxs})$, which is computationally more convenient, is determined as the sum of the wavelength-specific components $\mathrm{ln}(\mathcal{L}_{SZ})$ and $\mathrm{ln}(\mathcal{L}_X)$,
\begin{equation}
\begin{split}
    \mathrm{ln}\left(\mathcal{L}_{\jxs}\right) = & \: \mathrm{ln}\left(\mathcal{L}_{SZ}\right)+\mathrm{ln}\left(\mathcal{L}_X\right) \\
    = & -\frac{1}{2}\sum_{i=1}^{n_{SZ}}\left(\frac{f_i^{data_{SZ}}-f_i^{model_{SZ}}}{\sigma_i^{data_{SZ}}}\right)^2+ \\
    & +\sum_{j=1}^{n_X}\left[D_j\mathrm{ln}\left(M_j\right)-M_j-\mathrm{ln}\left(\Gamma\left(D_j+1\right)\right)\right].
\end{split}
    \label{eq:likelihood}
\end{equation}
The likelihood function for SZ data is the same as in \citet{Castagna2019}, where $f^{data_{SZ}}$ and $f^{model_{SZ}}$ are the observed and estimated surface brightness values, respectively, while $\sigma^{data_{SZ}}$ is the error measure, and $n_{SZ}$ represents the total number of available data points.
The likelihood function for the $n_X$ X-ray data \citep{Sanders2018} compares the predicted model values $M$ (including background) with the observed profiles $D$, assuming that the X-ray counts follow a Poisson distribution. $\Gamma$ is the usual gamma function. 

\subsection{Prior, posterior sampling, and diagnostics}
As presented in Sect.~\ref{subsec:parametrization}, \jxs\  allows extremely flexible modelling of the ICM, which involves a large number of parameters. Users have to select their prior distributions, and then the program estimates the posterior distribution thorugh an MCMC, marginalising over all of them according to the Bayesian approach. For the whole MCMC setup, we relied on the implementation refined by Sanders for \mb. The posterior is sampled with an affine-invariant ensemble sampler as proposed by \citet{Goodman2010} and implemented in \texttt{emcee} \citep{Foreman2013}.
The user has to specify the list of parameters to be fitted and optionally can change the prior distribution, whether uniform or Gaussian. The user is free to fix the desired number of random walkers, the number of iterations, and the burn-in period extent. The MCMC is initialised with default parameter values that can be also changed upon user request, and is automatically followed by a preliminary optimisation towards higher values of likelihood to facilitate convergence. Multi-threading computation is supported by \jxs \ and is strongly encouraged to minimise the execution time.

Qualitative and quantitative diagnostics (described in detail in Sect.~\ref{sec:ex}) are provided by \jxs\  to evaluate the convergence of the chains to the stationary distribution: traceplot and the cornerplot are automatically displayed, informing the user of the parameter evolution across iterations and of the joint posterior distribution; the acceptance fraction is reported in the program output; and plots with best-fit profiles and uncertainty intervals (at a probability level set at user choice) overplotted on the observed data are automatically produced.

Because the user can choose to fit different combinations of parameters, model selection plays an essential role in performing optimal analyses. When the models to be compared are nested (e.g. a fit with $\log(T_X/T_{SZ})=0$ compared to an analysis with free ratio), the Savage-Dickey density ratio is recommended \citep[e.g.][]{Trotta2007}. This index represents an approximation of the Bayes factor and can be computed as the ratio between the marginal posterior of the more complex model evaluated at the simpler model parameter value and the prior density of the more complex model evaluated at the same point.

\subsection{Execution-time balance}

The fit performed with \jxs\  can be slow, largely because of highly time-consuming operations included in the SZ processing, as was discussed in \citet{Castagna2019}.
The precise splitting of the CPU time across the SZ and X-ray parts of the analyses depends on the relative size of the fitted data, on the relative precision adopted for the Abel transform of the SZ and X-ray sides, and on whether the PSF convolution is performed on both sides or only on one. In our example below, we used a widely unbalanced setting: the X-ray side does not use any convolution by the point spread function (PSF), while the SZ convolution adopts an unnecessary small pixel to compute it and the Abel transform adopts again an excessively small pixel in the SZ side.
In such unbalanced conditions, the SZ fit accounted for 90\% of the CPU time, while the remaining 10\% was used for the X-ray fit. The fit can be made faster and less unbalanced between the two sides (e.g. 75\% vs. 25\%) without loosing accuracy by doubling the integration or convolution pixel size on the SZ side.
As always, it remains at user charge to determine the trade-off between execution time and the precision that satisfies their needs and requirements.

\subsection{Limitations}
To mention the main limitations of the program, \jxs\  analyses one object at a time (it cannot deal with superposed clusters or point sources; flagging should be used instead), assumes a centre for the cluster, spherical symmetry (information on the third dimension is mandatory for integrating along the line of sight), radius-independent metallicity (because joint X-SZ data sets with robustly measurable metallicity gradients are rare), and negligible covariance between errors of the different radial bins.

\section{A worked example} \label{sec:ex}
To highlight the capabilities of \jxs, we present an application of the program to real data. Our analysis purely illustrates the use of the code, it is not meant to be the most accurate astrophysical analysis of the cluster.

We analysed the high-redshift cluster of galaxies \clj\  ($z=0.89$), which has been variously studied by several authors in the past years in the X-ray \citep{Maughan2004, Maughan2007, Donahue2014} and in SZ \citep{ Mroczkowski2009, Korngut2011, Sayers2013, Adam2015, Romero2017, Romero2018}. \clj\  is a hot and massive cluster that was discovered in the WARPS survey \citep{Ebeling2001}. The cluster presents a relaxed morphology on a large scale, with some possible evidence of a disturbed core \citep{Maughan2007, Korngut2011, Adam2015, Romero2017}. We flagged the point source detected by \cite{Adam2015} at (RA, Dec) = (12:26:59.855,+33:32:35.21) with an aperture of 46".

\subsection{NIKA and \textit{Chandra} data}
The SZ observation, instrumental beam, transfer function, and the table of temperature-dependent Compton $y$ to mJy conversion factors used in the example come from the publicly available NIKA data release\footnote{http://lpsc.in2p3.fr/NIKA2LPSZ/nika2sz.release.php} \citep{Catalano2014}. We refer to \cite{Adam2015} for details about the cluster observation, whose reduction and filtering differ slightly, however.

We used \textit{Chandra} X-ray observations (OBSID=3180, PI. Ebeling, exposure time of 32 ks), reduced following the standard procedures using CIAO 4.1 and CALDB 4.5.2 \citep[e.g.][]{Andreon2019}. 
First, data were flare filtered. Then, point sources were detected by a wavelet detection algorithm and masked. Events in ten bands were extracted: [0.7-1], [1-1.3], [1.3-1.6], [1.6-2], [2-2.7], [2.7-3.4], [3.4-3.8], [3.8-4.3], [4.3-5.0], and [5-7] keV. Similar results were obtained by merging the central eight bands into just three.
We computed energy-dependent exposure maps to calculate the effective exposure time, accounting for dithering, vignetting, CCD defects, gaps, and flagged pixels.  We then measured counts and effective exposure time in the ten bands in circular annuli with increasing width with radius to counterbalance the decreasing intensity of the cluster. The minimum width is taken to be 3 arcsec, which is larger than the \textit{Chandra} PSF.
We only considered radii where the exposure time was longer than 50\% of the on-axis exposure time and annuli included in the field of view by more than two-thirds. 
The cluster centre was iteratively computed as the centroid of X-ray emission within the inner 20 kpc, and the same centre was adopted for the SZ data.
As background, we used blank field images \citep[using CIAO \textsc{blanksky},][]{Fruscione2006}, normalised to the count rate in the hard band [9-13] keV, and we derived the background surface brightness in the ten bands accounting for exposure time variations, dithering, vignetting, CCD defects, gaps, and flagged pixels. 
In the X-ray data, information about the temperature profile is encoded in the ratio of the cluster count-rate profiles, while much of the metallicity information is contained in the [3.4-3.8] keV band.

As a proof of the capabilites of \jxs, we used data with different radial sampling between SZ and X-ray, the latter sampled on an
irregular grid. Because the model is integrated in the same bins of the observations, results do not depend on
binning (except for unreasonable choices, such as having one single bin, which completely loses spatial information).

\begin{figure}
    \centering
    \includegraphics[page=1, width=\linewidth]{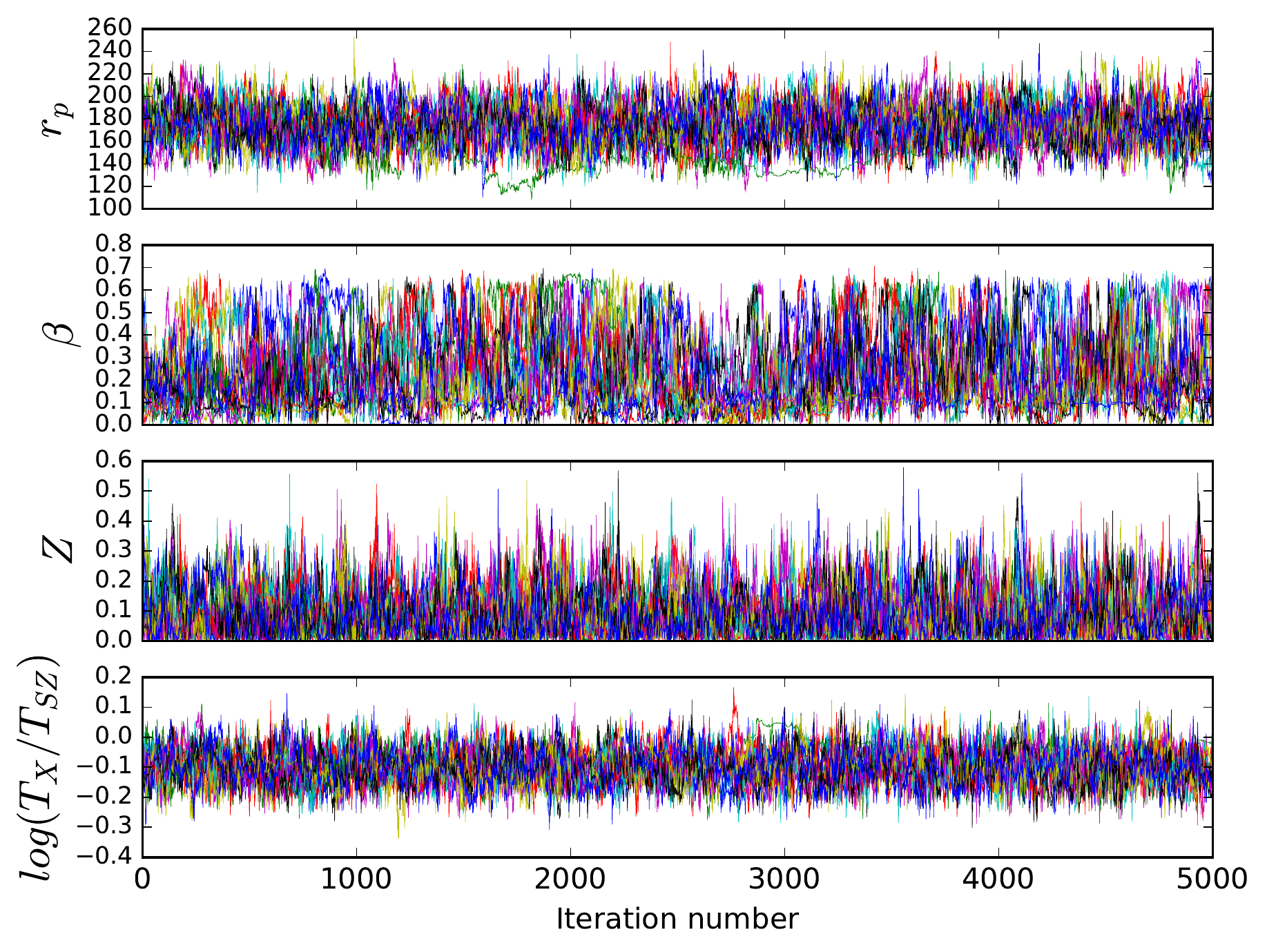}
    \caption{Trace plot automatically produced by \jxs. Trace plots for four variables are only shown here, but are automatically produced for all parameters.}
    \label{fig:diagnostics}
\end{figure}

\begin{figure*}[b]
    \centering
    \includegraphics[width=\linewidth]{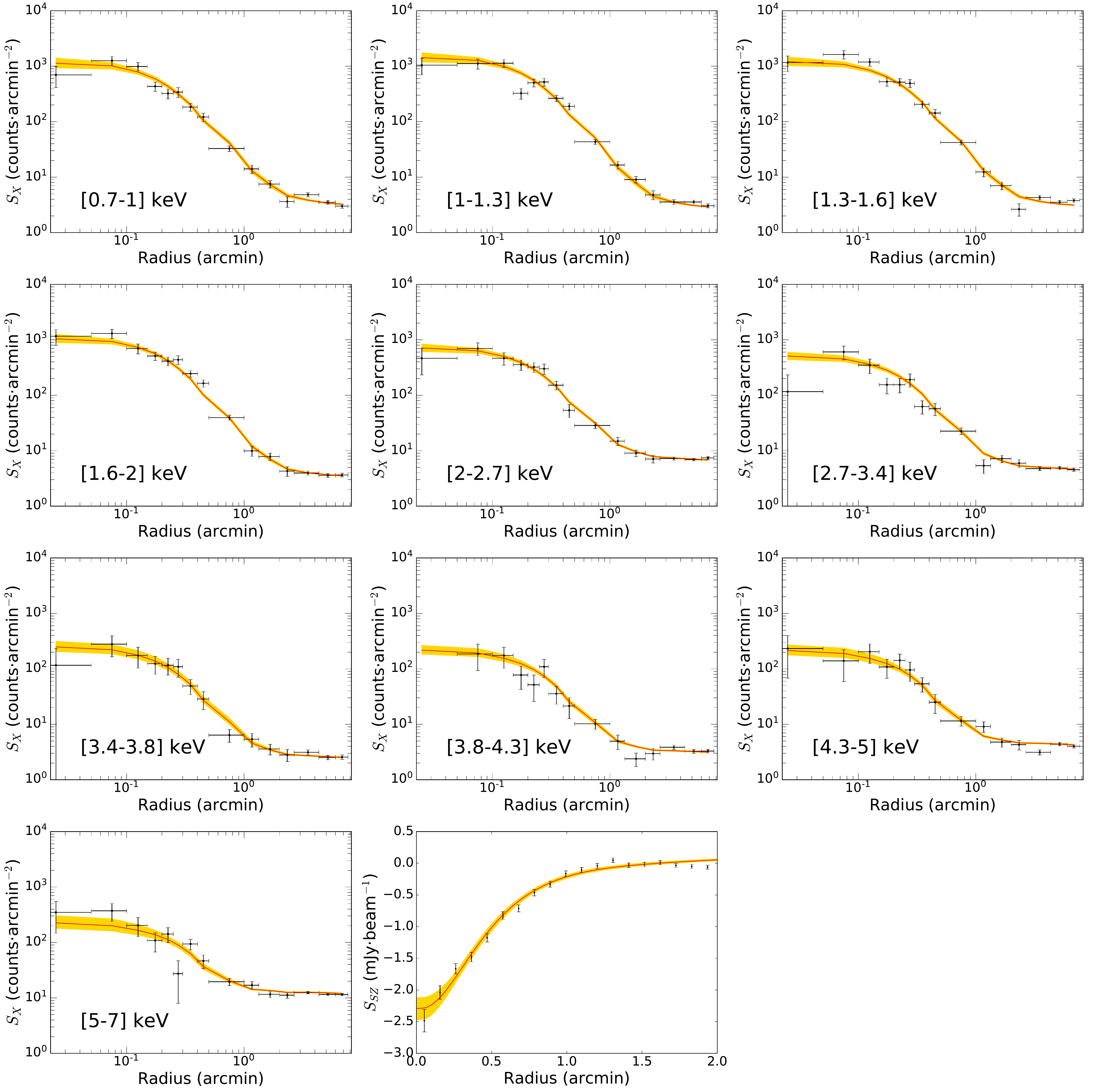}
    \caption{X-ray and SZ projected radial surface brightness profiles automatically produced by \jxs. Red lines show the best-fit  profile. Shaded yellow areas represent 95\% intervals. Blue points show X-ray data. Black points show SZ data. The X-ray profiles flatten off to the background value.}
    \label{fig:fitall}
\end{figure*}

\begin{figure*}[t]
    \centering
    \includegraphics[width=\linewidth]{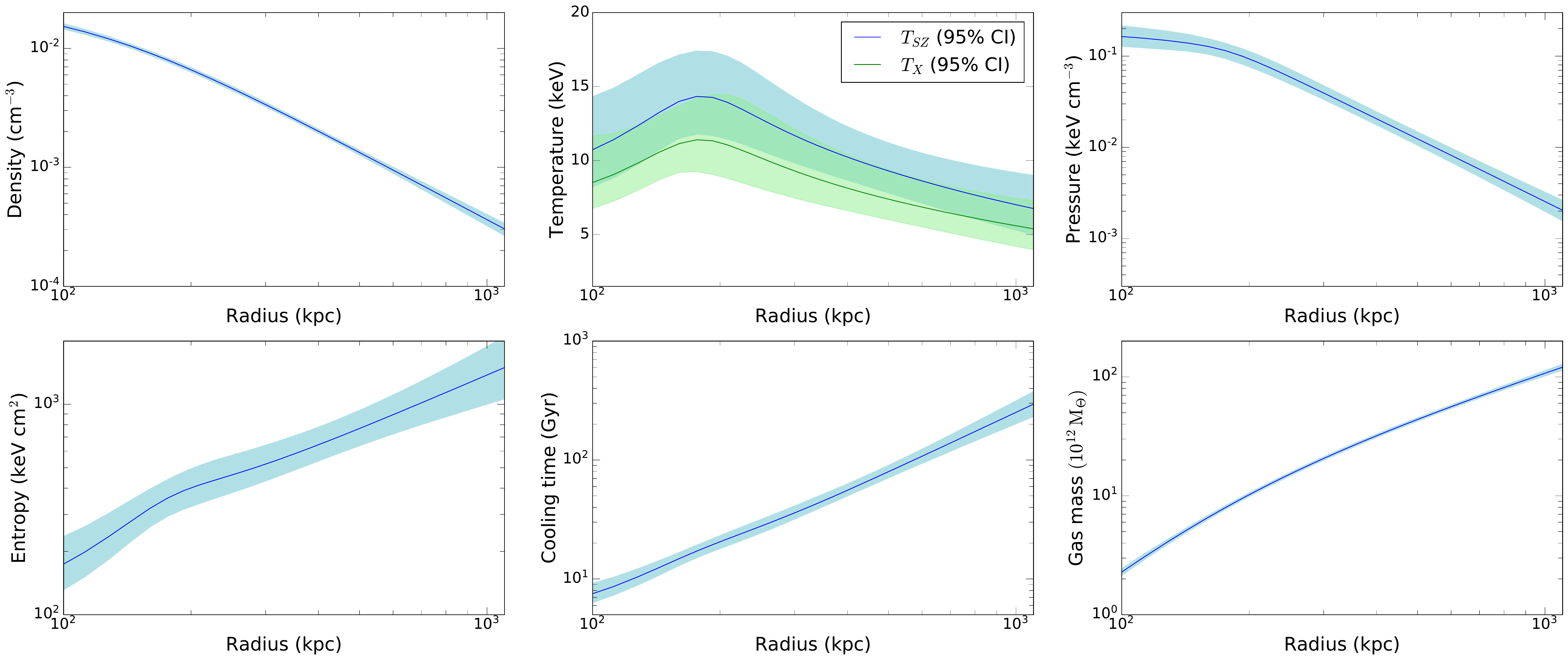}
    \caption{Deprojected radial profiles for the main thermodynamic properties of the cluster (median with 95\% intervals).}
    \label{fig:vertprofs}
\end{figure*}
\subsection{Model definition}
As defined by the equations in Sect.~\ref{subsec:parametrization}, many parameters are involved in the \jxs\  fit: the gNFW pressure profile has five parameters ($P_0, r_p, a, b, \text{and } c$), the Vikhlinin density profile has 10 ($n_{e0}, r_c, \alpha, \beta, r_s, \gamma, \epsilon, n_{e02}, r_{c2},\text{ and } \beta_2$), to which need to be added the metallicity $Z$; the temperature profile ratio $\log(T_X/T_{SZ})$; the parameter that accounts for the calibration uncertainty of the SZ measurement, implemented as a multiplicative component on the conversion; and the backscale parameter, which controls the scaling of the X-ray background. This yields a total maximum number of 19 parameters that can be fitted at the same time.

For our example, we cancelled out the additive component of the electron density profile by fixing $n_{e02}=0$, we assumed $\alpha=0$ and $\gamma=3$ following the reasoning of \citet{Mroczkowski2009, Comis2011, Adam2015}, and we fixed $c=0.3081$, in accordance with the universal pressure profile from \citet{Arnaud2010}. 
In addition to these parameters, we accounted for the calibration and the backscale. We alternately considered the logarithm of the temperature profile ratio $\log(T_X/T_{SZ})$ fixed to 0 or let free to vary, which means that we considered 13 parameters at maximum.

We took a Gaussian prior with $\sigma=0.07$ centred on $1$ for the SZ calibration \citep{Adam2015}, a Gaussian prior with $\sigma=0.1$ centred on $1$ for the backscale parameter, and uniform priors for all the remaining parameters, with ranges $0<P_0<1$, $100<r_p<1000$, $0.5<a<20$, $0.5<b<10$, $0.0001<n_{e0}<1$, $0.1<r_c<r_s<2.5$, $0<\beta<4$, $0<\epsilon<10$, $0<Z<1$, and $-1~<~\log(T_X/T_{SZ})~<~1$, where $P_0$ is expressed in keV cm$^{-3}$, $n_{e0}$ in cm$^{-3}$, and $r_p$, $r_c$, $r_s$ in kpc.

We ran 70000 iterations of 30 walkers and considered the first 20000 as the burn-in period. In order to reduce autocorrelation, we stored $N/100$ samples for each walker.

We considered an SZ pixel size of 2 arcsec (the PSF is $\sim$18 arcsec) and a cluster radial extent of 5 Mpc for the Abel integral computation. We fixed the absorbing column at the Galactic value \citep{Kalberla2005}. We adopted a flat $\Lambda$CDM cosmology with $H_0=67.32 \mathrm{\;km\;s^{-1}\;Mpc^{-1}}$, $\Omega_M=0.3158,$ and $\Omega_\Lambda=0.6842$ \citep{Planck2018}.

\subsection{Results}
We fitted the data several times to understand the effect of our assumptions on the derived thermodynamic profiles. We also compared \jxs \ performances with a restricted joint analysis and an X-ray alone analysis.

\subsubsection{Disjointed temperatures $T_X$ and $T_{SZ}$ with or without HE} \label{subsec:fit1}
Our reference analysis assumes HE, allows X-ray and SZ temperatures to differ from each other, and has 13 free parameters (4 of which for the pressure profile, and 5 for the electron density profile).
Figure~\ref{fig:diagnostics} and Fig.~\ref{fig:corner}, automatically produced by \jxs, show the trace plot and joint plus marginal posterior distribution, respectively. They inform about the convergence of the chains and the parameter estimates.
The acceptance fraction is found to be 0.11, which is lower than recommended; nevertheless, the wide thinning we adopted allowed us to have sufficiently uncorrelated samples. Some expected degeneracies are evident in the joint distributions: $\beta$ and $\epsilon$ are highly negatively correlated, $r_c$ and $\beta$ are positively correlated, as are $r_p$ and $b$, while $n_{e0}$ is negatively correlated with $r_c$. The $a$ parameter is not well constrained by the data we considered.
Figure~\ref{fig:fitall}, which is automatically generated as well, shows the best-fit profiles and their 95\% uncertainty (equal-tailed credible interval) on top of the fitted data for each of the X-ray and SZ bands.
Figure~\ref{fig:vertprofs}, also automatically produced, shows the main radial thermodynamic profiles of the cluster (median values and 95\% intervals). 
The estimated entropy profile is in line with the \citet{Voit2005b} fit to non-radiative simulations, as adapted by \citet{Pratt2010}.

\begin{figure}[b!]
    \centering
    \includegraphics[width=\linewidth]{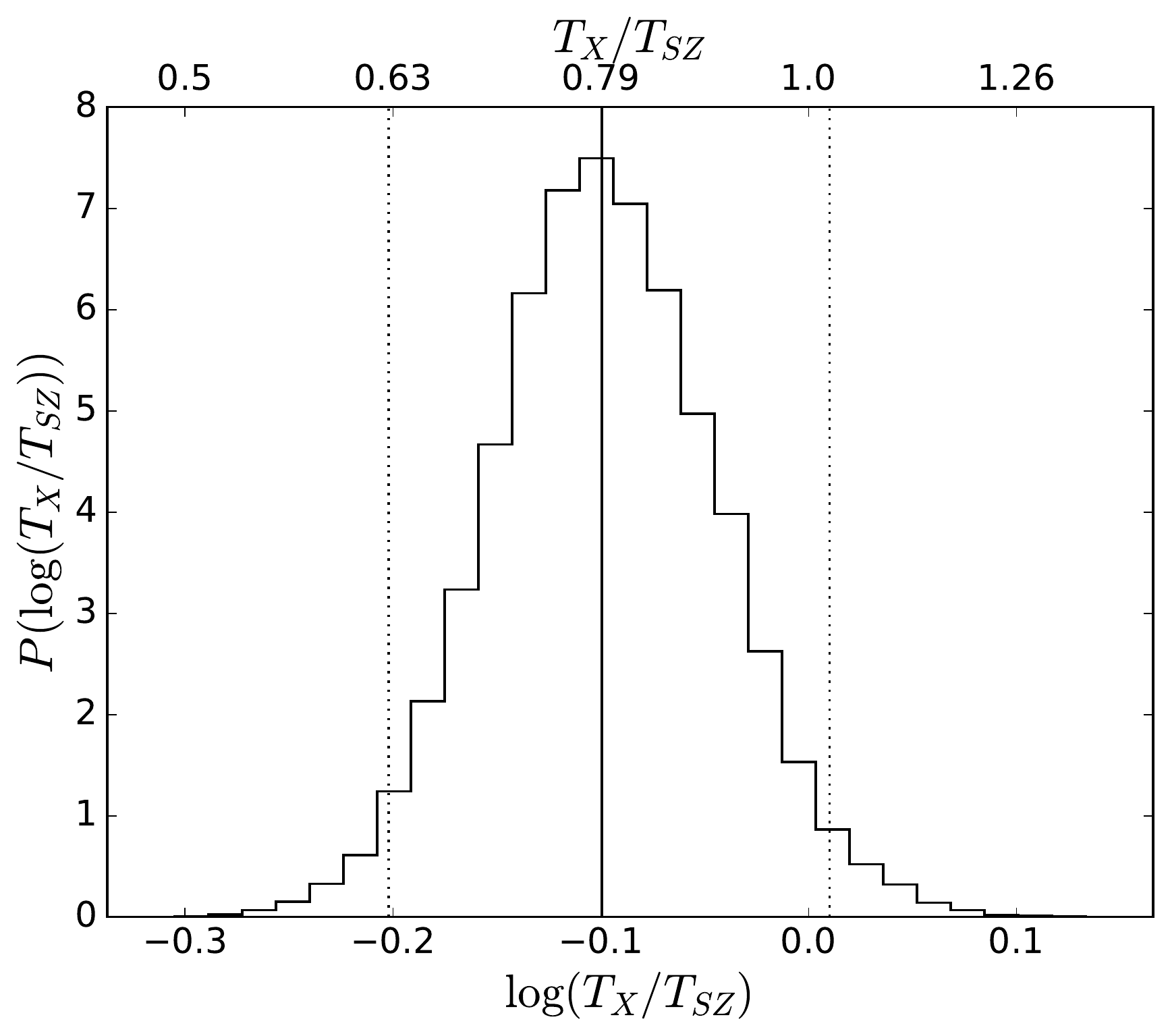}
    \caption{Posterior distribution of the temperature profile ratio parameter $\log(T_X/T_{SZ})$. Median value and 95\% intervals are highlighted.}
     \label{fig:tratio}
\end{figure}
\begin{figure}[b!]
    \centering
    \includegraphics[width=\linewidth]{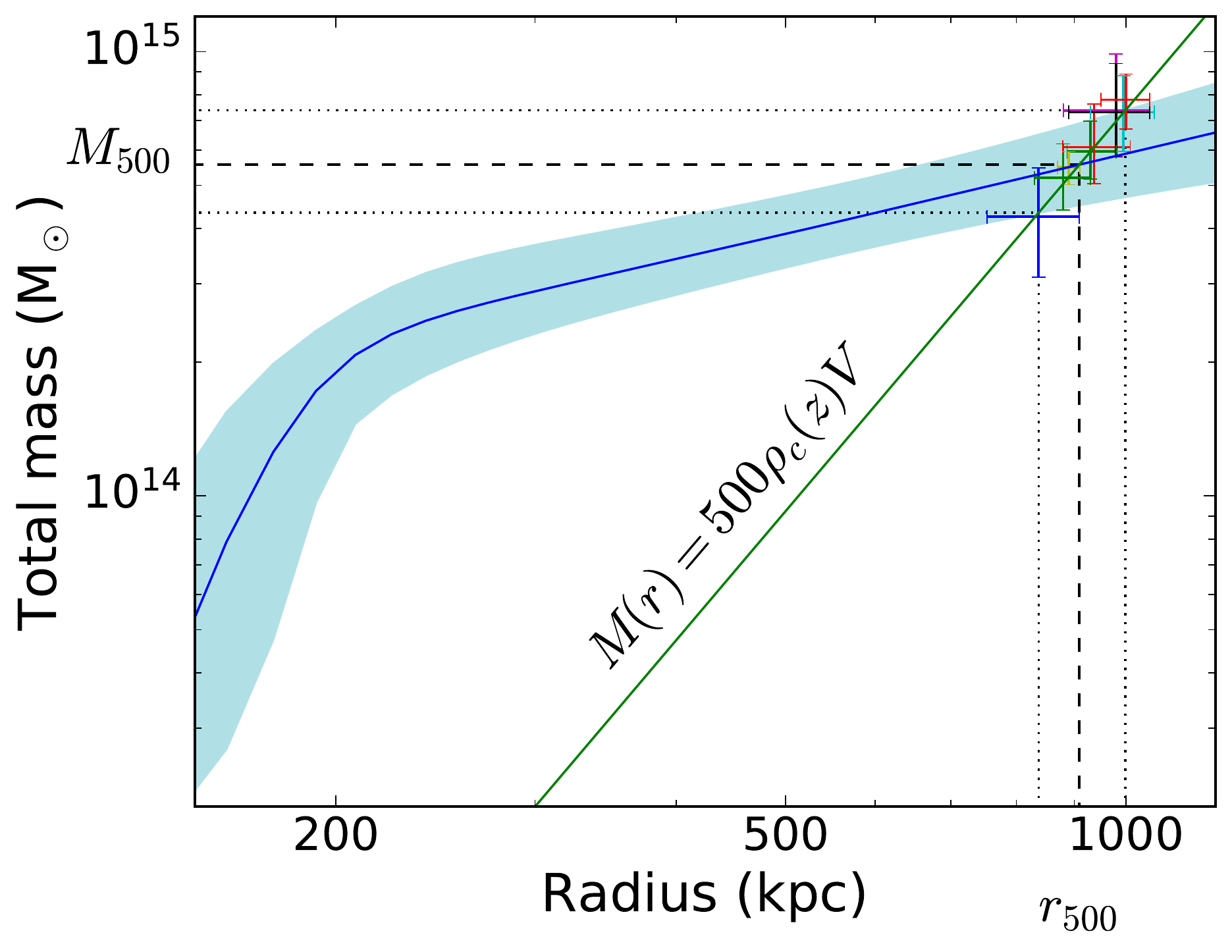}
    \caption{Mass profile. The blue line shows the median mass profile derived assuming hydrostatic equilibrium (95\% intervals shaded). Points with error bars represent $r_{500}$ and $M_{500}$ from the literature (references in the text).
    }
    \label{fig:mass_lit}
\end{figure}
\begin{figure*}[htp]
    \centering
    \includegraphics[width=\linewidth]{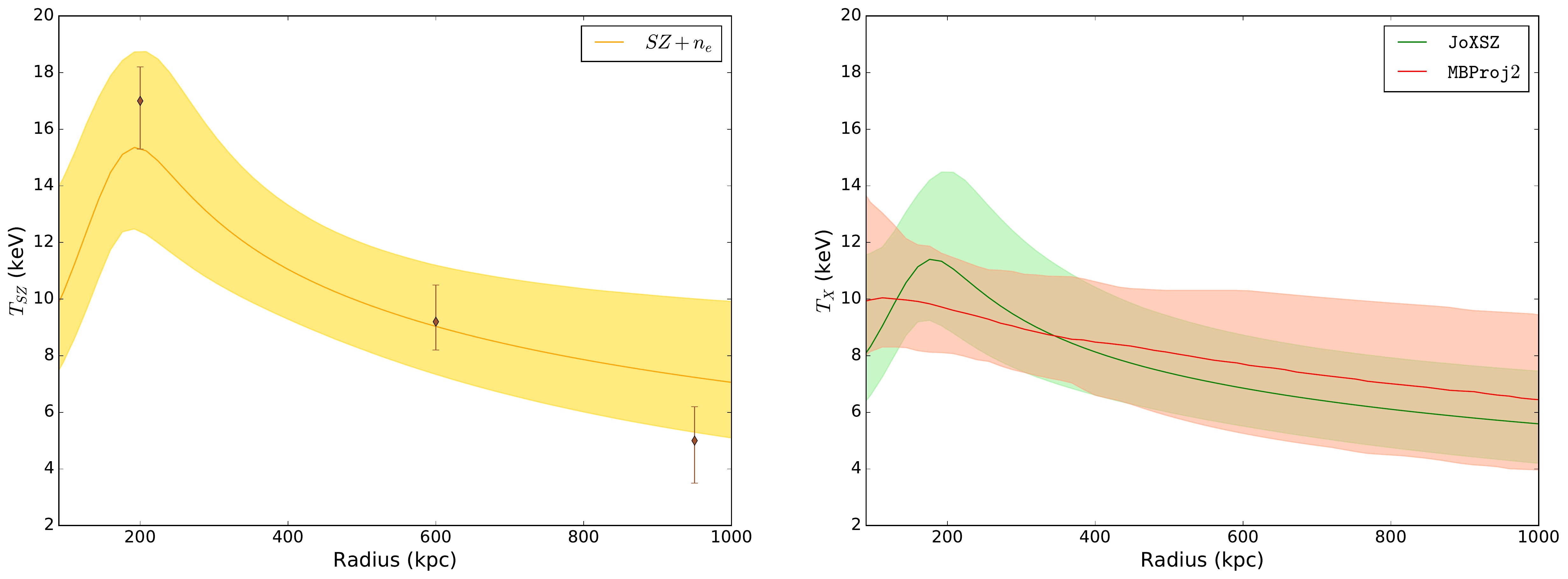}
    \caption{Comparison of \clj\  deprojected $T_{SZ}$ (left panel) and $T_X$ (right panel) temperature profiles (median with 95\% interval). The restricted joint analysis is shown in yellow-orange. Green shows \jxs. Red shows the X-ray analysis alone. The posterior derived by \citet{Adam2015} is marked by points.}
    \label{fig:temp}
\end{figure*}

\begin{figure}[b!]
    \centering
    \includegraphics[width=\linewidth]{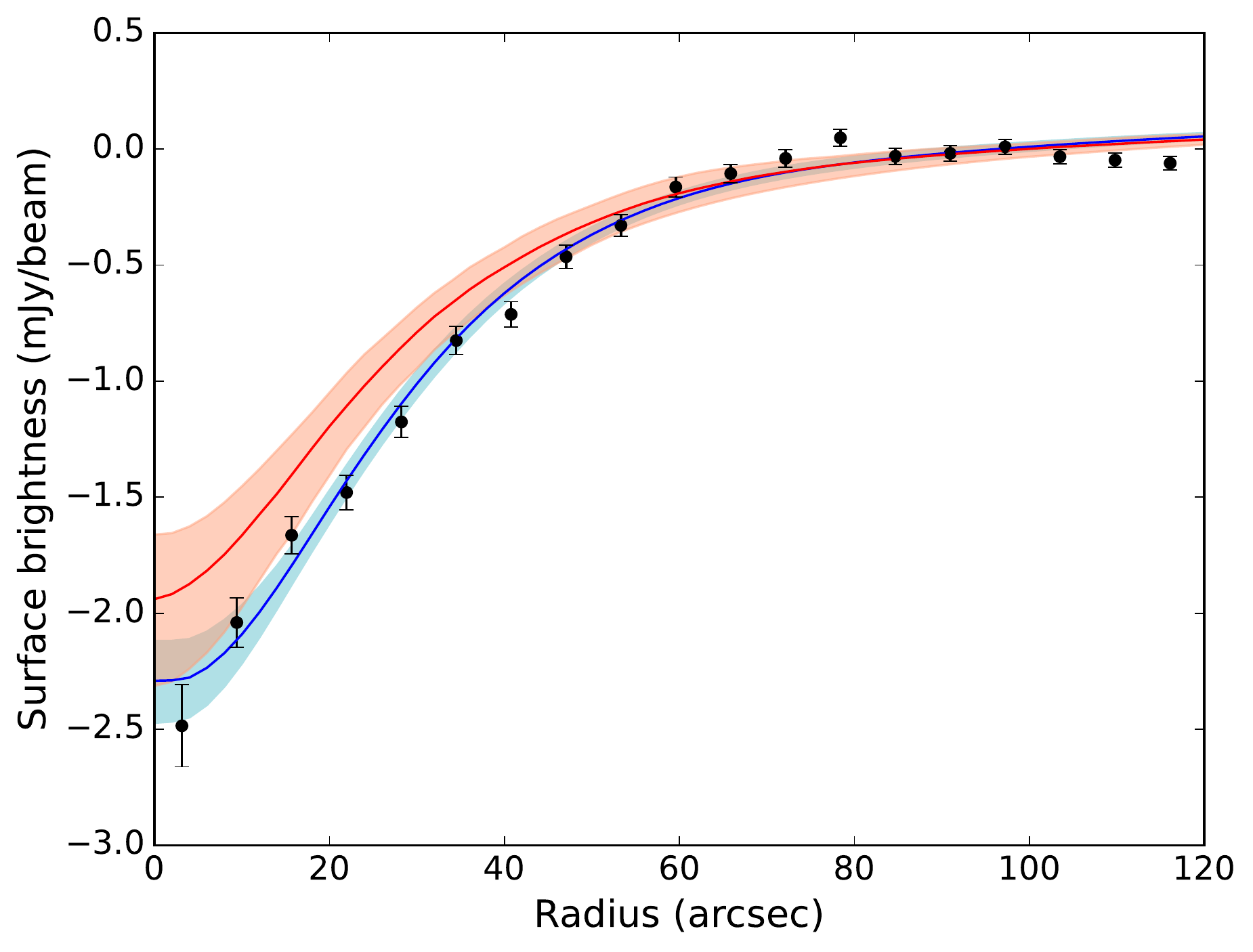}
    \caption{Comparison of \clj\  projected SZ surface brightness profiles (median with 95\% interval). Blue shows \jxs. Red presents predictions from the X-ray alone analysis.
    Points with 68\% error bars show the SZ data from NIKA.}
     \label{fig:fit_sz_data}
\end{figure}

Figure~\ref{fig:vertprofs} shows the X-ray and SZ temperature profiles.
The parameter $\log(T_X/T_{SZ})$ is found to be $-0.10^{+0.11}_{-0.10}$, and its posterior distribution is shown in Fig.~\ref{fig:tratio}. The ratio between the X-ray and SZ temperatures has an unusual value, it is particularly far from 1 compared to what is expected from clumping effects as estimated from numerical expectations \citep{Nagai2011}. This might be related to calibration systematics, such as the long-standing \textit{Chandra}-XMM $T$ systematics \citep{Schellenberger2015}, or possible inaccuracies of the transfer function \citep{Romero2019}. 
We emphasise that the conclusions accuracy does not depend exclusively on the correctness of operations that are performed on the data (the fitting code), but also on the lack of systematics in the input quantities (data, PSF, transfer function, instrument calibrations, etc.).

In absence of systematics, to estimate whether the data provide evidence in favour of differences between the X-ray and SZ temperatures, users may compute the Bayes factor from the \jxs \ output using the Savage-Dickey ratio, that is, the ratio between the posterior and prior density probabilities at $\log(T_X/T_{SZ})=0$. In our case, we found a Bayes factor of 3.1 ($=1.53/(1/2)$), which is almost inconclusive. 

Figure~\ref{fig:mass_lit} shows the mass profile estimated assuming HE and its 95\% uncertainty. The intercept with $500\rho_c(z)V$ gives $r_{500}=910^{+89}_{-72}$ ~kpc and $M_{500}=5.57^{+1.81}_{-1.23}\times 10^{14}$~M$_\odot$, consistent with the measures obtained by \citet{Maughan2007}, \citet{Mroczkowski2009}, \citet{Mantz2010}, and \citet{Adam2015}. 
The profile shape is in accordance with the shape obtained by \citet{Adam2015}.
Relaxing the assumption of HE (Sect.~\ref{subsec:mass}) has barely any effect on our results: we performed a separate analysis without assuming a monotonically increasing mass profile, and only 0.1\% of the samples returned unphysical values, likely because the parameters are sufficiently well constrained to work without this additional prior.

\subsubsection{Restricted joint SZ+$n_e$ and X-ray alone} \label{subsec:comparison}
In a first fit, a restricted joint analysis SZ+$n_e$, we fixed the metallicity to $Z=0.3$ solar \citep{Maughan2007, Arnaud2010}, we discarded all X-ray photons except for those in the [0.7-1] keV band, but kept HE and $\log(T_X/T_{SZ})=0$.
This is meant to mimic some standard SZ+$n_e$ fits mentioned in Sect.~\ref{sec:intro} \citep[for an exception, see][]{Romero2017}.
Unlike most analyses, we used the correct Poisson-Gauss expression for the likelihood and a more inclusive error budget, and we fully propagated errors on the derived thermodynamic profiles.

The left panel of Figure~\ref{fig:temp} shows our SZ gas-mass weighted posterior temperature profile.
Because we discarded the information in the X-ray spectrum, the temperature derived here is between 1.1 and 1.3 times more uncertain than in the fully joint fit (i.e. more uncertain than in Fig.~\ref{fig:vertprofs}) for all $r~>~200$ kpc. 
\citet{Adam2015} performed a restricted joint analysis of SZ data (of much the same NIKA data, supplemented by \textit{Planck}) and electron density profile based on \textit{Chandra} X-ray data.
Our restricted joint SZ+$n_e$ posterior temperature profile agrees with theirs (points).

The second analysis completely excluded the SZ data and was performed with the original \mb \ code.
In particular, we used the same electron density parametrisation as adopted in \jxs\  and a simplified \citet{Vikhlinin2006} temperature profile, as in \citet{McDonald2014}.
The right panel of Figure~\ref{fig:temp} compares the X-ray temperature profile derived in this way with the profile derived in our reference fully joint fit.
The two profiles largely overlap because the free $\log(T_X/T_{SZ})$ reduces the information passage between the SZ and X-ray parts of the analysis, and allows the \jxs\  X-ray temperature to agree with the temperature derived from X-ray data alone.

Figure~\ref{fig:fit_sz_data} explains why the X-ray temperature is systematically lower than the SZ temperature, as highlighted in Fig.~\ref{fig:vertprofs}. The observed SZ surface brightness profile is higher in absolute value than the profile predicted based on the X-ray temperature and density profiles alone.

\section{Conclusion} \label{sec:conc} 
The recent spread of SZ observations, together with the vast availability of X-ray data, caused joint SZ+X analyses of galaxy clusters to flourish. The need for a public tool to perform such analyses was shared among the community  of workers within the field.
We presented \jxs, the first publicly available code that combines the multiwavelength SZ+X approach and the X-ray multiband fitting technique in a full and consistent way.
\jxs \ supports data coming from multiple instruments and can even handle missing data for either SZ or X-ray surface brightness, follows a Bayesian forward-modelling approach, and adopts flexible parametrisations of the thermodynamic cluster profiles.
It makes full use of the information contained in the observations, it uses the correct Poisson-Gauss expression for the joint likelihood, accounts for beam smearing and transfer function, includes SZ calibration and X-ray background systematics, adopts a consistent temperature in the various parts of the code, and allows differences between SZ and X-ray temperatures, for example if users wish to study cluster elongation or clumping. By also fitting the cluster metallicity, \jxs\  accounts for the metallicity dependence of the X-ray conversion factor.
It supports the use of different radial binnings for SZ data and X-ray data and consents either to adopting or to relaxing the assumption of HE on user request.
For these reasons, \jxs\  goes beyond simple SZ and electron density fits.
When HE holds, \jxs\  uses a physical (positive) prior on the derivative of the mass profile and derives the mass profile and overdensity radii $r_\Delta$.
As in other approaches, \jxs\  makes the usual assumptions about cluster sphericity, and when requested, HE.

\jxs\  returns convergence diagnostics, parameter estimates with uncertainties, joint and marginal probability distributions, projected radial distributions, and three-dimensional thermodynamic radial profiles.

The code is fully documented, and the users are free to customise their analysis in accordance with their needs and requirements.
\jxs\  has been released as an open-source Python project, and its code is publicly available on 
GitHub.

We also provided an application using real data of the high-redshift galaxy cluster \clj\  to illustrate the features of \jxs.
When compared to a restricted joint SZ+$n_e$ fit or an X-ray alone fit, \jxs\  derives smaller uncertainties by accounting for the whole information of SZ and multiband X-ray data. While the uncertainty reduction is not striking for our case, a more relevant improvement may occur in different circumstances.
The easiness with which thermodynamic profiles are derived with \jxs\  allows the users to focus on astronomically interesting features.

We plan to further develop the code in the near future to also fit the shear (i.e. weak-lensing information). The execution time of the SZ part of the analysis will also be improved.

\begin{acknowledgements}
We warmly thank Charles Romero, Roberto Scaramella and Luca Di Mascolo for their useful comments on the manuscript. We are also thankful to the anonymous referee for the detailed revision that helped improve the quality of the paper.
F.C. acknowledges financial contribution from the agreement ASI-INAF n.2017-14-H.0 and PRIN MIUR 2015 Cosmology and Fundamental Physics: Illuminating the Dark Universe with Euclid.
\end{acknowledgements}

\bibliographystyle{aa}
\bibliography{references}

\begin{appendix}
\section{Creation of Python classes \textit{pressure} and \textit{SZ data} and joint likelihood definition} \label{app:py}
The \jxs\  code is grafted onto the structure designed for the \mb\  code by \citet{Sanders2018}. 
To fit X-ray surface brightness profiles alone, \mb\  allows users to model one or more thermodynamic profiles except for the pressure profile, and it does not support any fit to SZ data.

The novelty we introduced in \jxs\  is the inclusion and treatment of the SZ data and the possibility of parametrising the pressure profile by creating a specific class for it. The functions within the class replicate the standard functions for the other thermodynamic profiles: \texttt{defPars} defines the default parameter values, \texttt{press\_fun} computes the analytic expression of the pressure profile, and \texttt{press\_derivative} returns the analytically computed first derivative of it. Similarly, a new class for the temperature is defined as the ratio between pressure and electron density.

We also created a specific Python class to recollect all the elements required to perform the fit on SZ data that do not change in the iterations: the sampling step and the corresponding radius vector, the matrices of beam and transfer function built from observed or approximated data, the conversion factors from Compton $y$ to surface brightness, and the observed SZ data.

As outlined in Fig.~\ref{fig:flow} and described in Sect.~\ref{subsec:model}, the likelihood function $\mathcal{L}_{\jxs}$ is defined as the product of $\mathcal{L}_X$ from \mb \ and $\mathcal{L}_{SZ}$ from \ppf, the former updated for ignoring missing values and the latter updated to allow the temperature dependence of the Compton $y$ to surface brightness conversion.
Prior to the likelihood calculation, if the unphysical mass exclusion is set, the mass profile is evaluated according to Eq.~\ref{eq:HE} and $\mathcal{L}_{\jxs}$ is properly set to $-\infty$.

\section{Joint and marginal posterior probability distributions}
\begin{figure*}
    \centering
    \includegraphics[width=\linewidth]{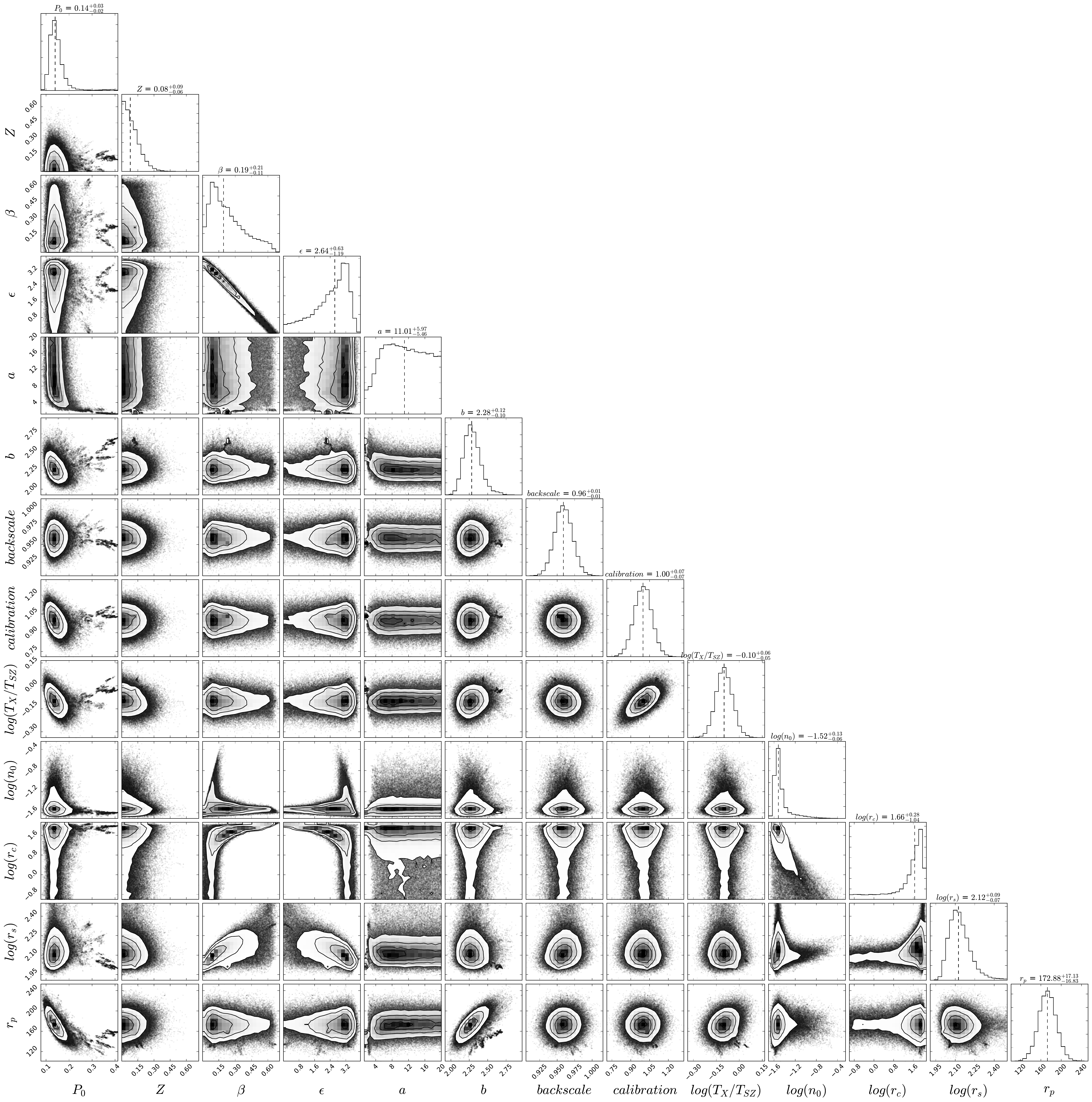}
    \caption{Joint and marginal posterior distributions that are automatically produced by \jxs. Dashed lines in the diagonal panels indicate median values.}
    \label{fig:corner}
\end{figure*}
\end{appendix}

\end{document}